\documentstyle[12pt,psfig]{article}
\begin{document}
\begin{center}
{\Large \bf Evidence of Pentaquark States from
$K^+ N$ Scattering Data?}
\vskip0.5cm
N. G. Kelkar$^{1}$, 
M. Nowakowski$^{1}$ and K. P. Khemchandani$^2$ \\
{$^1$\it Departamento de Fisica, Universidad de los Andes, 
Cra. 1E No.18A-10, Santafe de Bogota, Colombia}\\
{$^2$\it Nuclear Physics Division, Bhabha Atomic Research Centre, \\
Mumbai 400 085, India} 
\end{center}
\begin{abstract}
Motivated by the recent experimental evidence of the exotic $B = S = +1$ 
baryonic state $\Theta$ (1540), we examine the older existing data 
on $K^+ N$ elastic scattering through the time delay method. 
We find positive peaks in time delay around $1.545$ and $1.6$ GeV in 
the $D_{03}$ and $P_{01}$ partial waves of $K^+ N$ scattering respectively, 
in agreement with experiments. We also find an indication 
of the $J=3/2$ $\Theta^*$ spin-orbit partner to the $\Theta$, in the $P_{03}$ 
partial wave at 1.6 GeV. 
We discuss the pros and contras of these findings in support
of the interpretation of these peaks as possible exotics.   
\end{abstract}
\vskip0.5cm
In a recent letter by Nakano {\it et al.} \cite{nakano}, strong 
experimental evidence 
for the exotic state, characterized by the quantum numbers
$B = S = +1$, with a mass and width of $1.54\pm0.01$ and $0.025$ GeV 
respectively, was reported. This could be interpreted as a molecular
meson-baryon resonance 
(see e.g. \cite{barnes}) or a pentaquark baryon \cite{jaffe,newpenta,gloz}, 
hence its importance. It was actually since the late sixties that 
such a state was predicted \cite{roy}.   
Some other groups \cite{clas} also reported similar results which shifts the
evidence for the new state from plausibility to certainty. In this note,
we would like to shed some light on these findings 
by examining whether this exotic
state around 1.54 GeV is
actually supported by phase shifts \cite{arndt92} obtained from older 
data on kaon-nucleon elastic scattering. 
If our analysis is taken in support of the experimental interpretation, 
we can assign or speculate
about the $L_{2I,2J}$ quantum numbers. Indeed, according to our analysis, 
the exotic state found in \cite{nakano} and \cite{clas} is likely to be either
a $P_{01}$ (this would be in agreement with the prediction made in
\cite{diak} which actually prompted the experiments) or $D_{03}$
(or maybe even $P_{03}$).

In \cite{we1}, we used the time delay method of Eisenbud, Wigner
\cite{wigner}
and Smith \cite{smith} (see also \cite{brans,peres,rest})
to detect resonances from the phase shift derivative.
The peak in the time delay given by,
\begin{equation}\label{1}
\Delta t(E) = 2 \hbar {d\delta_l(E) \over dE}
\end{equation}
where $\delta_l$ is the phase shift in the $l^{th}$ partial wave, 
corresponds to the resonance position $E=E_R$. 
In \cite{we1}, we found that this method yields  
an excellent agreement with the established meson resonances. 
We noticed that the time delay method, which is
the quantification of phase shift motion (in the extreme one says that the
phase shift makes a $180^0$ jump passing through $90^0$),
is very sensitive to `small structures' in data, which might be overlooked 
otherwise.

The time delay in scattering can also be expressed in terms of the
transition matrix as \cite{we1},
\begin{equation}\label{4}
S^*_{ij} \,S_{ij}\, \Delta t_{ij}\, =\, 2 \,\hbar\,
\biggl[ \Re e \biggl({dT_{ij} \over dE}\biggr)\,+ \,2 \,\Re e T_{ij}\,\,
\Im m \biggl ({dT_{ij} \over dE}\biggr) \,-\, 2\, \Im m T_{ij}\,\,
\Re e\biggl( {dT_{ij} \over dE}\biggr)\, \biggr],
\end{equation}
where $i$ and $j$ represent the incident and outgoing channels in
a scattering process. 
In \cite{we2}, we applied the time delay method to find
resonances in $K^+ N$ elastic scattering, using the above relation and model
dependent $T$-matrix solutions \cite{arndt92}. 
We found evidence for low-lying $K^+ N$ resonances around 1.5 GeV in
the $P_{01}(1.57)$, $P_{13}(1.48)$ and $D_{03}(1.49)$ partial waves. 
It is important to note that analyses of the
$K^+ N$ elastic scattering data using other methods such as 
poles or argand diagrams have not found resonances around 1.5 GeV.
In the present work we do not use the model dependent solutions of the
$T$-matrix along with eq. (2), but
rather the single energy values of phase shifts to evaluate time delay
using eq. (1). We use these phase shifts to reduce the model dependence
in our calculations and compare our results obtained from the older
data with the recent experimental findings.

Before going into details of the analysis of the exotic case, let us
see how the method of time delay works in the case of established
$N^*$ resonances. This is done in Fig. 1, for the standard 
($B = 1$, $S = 0$) $D_{13}$ resonance.
\begin{figure}[h]
\centerline{\vbox{
\psfig{file=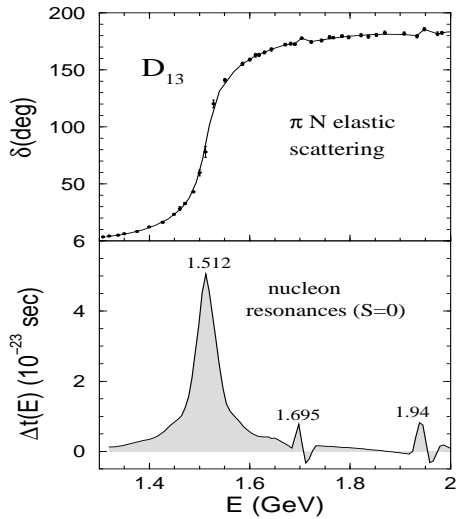,height=7cm,width=6cm}}}
\caption{Single energy values of phase shifts (filled circles) in the
$D_{13}$ partial wave of $\pi N$ elastic scattering. 
The time delay evaluated from the fit (solid line) to phase shifts, 
displaying the known nucleon resonances is shown in the lower half of
the figure.}  
\end{figure}
We see a prominent peak of the four-star $D_{13}(1520)$, but also smaller
peaks for the three-star $D_{13}(1700)$ and the less established two-star
$D_{13}(2080)$ resonances. The smaller peaks are due to the small phase 
shift motion which we took into account by 
making a polynomial fit to the single energy values, 
rather than using the energy dependent $T$-matrix 
solutions. Though in principle, such a fit 
is not always the best way of analysing data, it is justified here by
the fact that (i) even small phase shift motion seems to agree with
established resonances and (ii) we think that a $\chi^2$ fit would 
give the same results given the very small error bars (i.e. even the 
smaller peaks will not vanish) (iii) in \cite{we2} we evaluated time
delay using smooth model dependent solutions of the $T$-matrix and 
obtained very similar results (iv) in \cite{we1} we applied the same
method of fitting single energy values and confirmed all existing
meson resonances (well established and less established) (v) last but
not least the peak values found in the present work are very close
to those found by recent experiments. It is also interesting
to note that a peak in the speed plots was found at 1540 MeV in
the $P_{01}$ partial wave in an old $K^+N$ scattering experiment 
\cite{nakaji}.  
Note also that calculating the time delay from single energy values of
phase shift is less model dependent than the model dependent solutions.
\begin{figure}
\centerline{\vbox{
\psfig{file=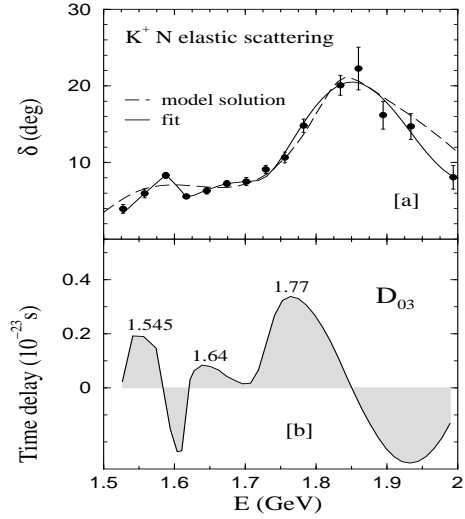,height=7cm,width=6cm}}}
\caption{
(a) Single energy values of phase shifts (filled circles), model dependent
solutions (dashed line) [3] and fit (solid line) (b) time delay in 
$K^+N$ elastic scattering evaluated from the fit to the phase shift 
(solid line).} 
\end{figure}
\begin{figure}
\centerline{\vbox{
\psfig{file=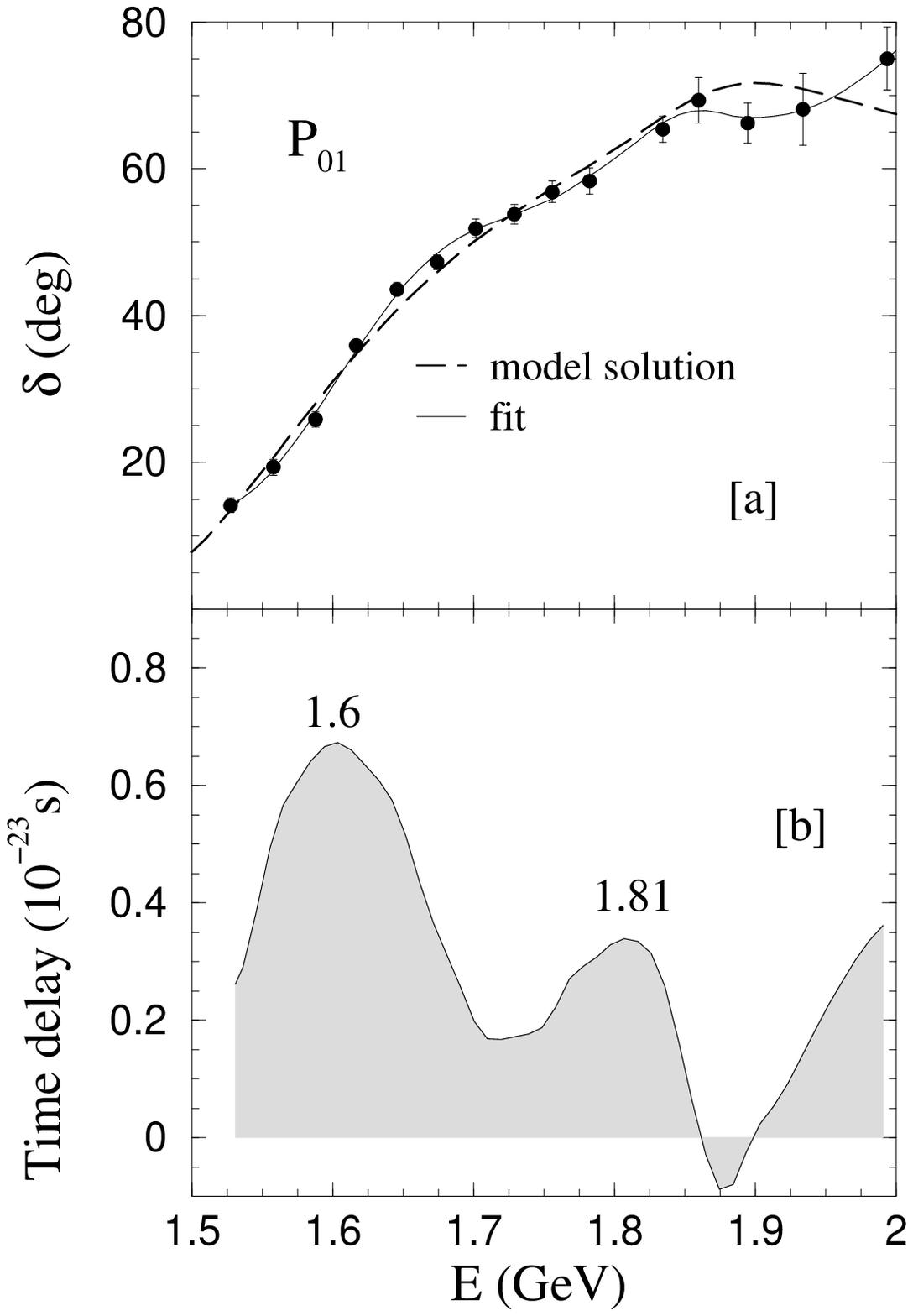,height=7cm,width=6cm}}}
\caption{
Same as Fig. 2}
\end{figure}
\begin{figure}
\centerline{\vbox{
\psfig{file=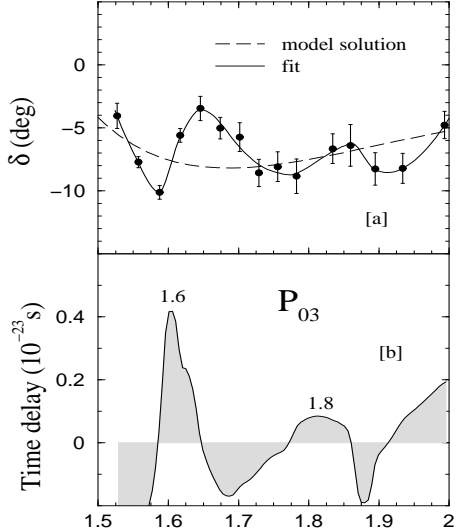,height=7cm,width=6cm}}}
\caption{
Same as Fig. 2}
\end{figure}

Motivated by the recent evidences \cite{nakano,clas}, 
we decided to evaluate the time delay (as in eq. (\ref{1})) 
using the available 
$K^+ N$ scattering phase shifts and check for small structures
which could give rise to time delay peaks. 
In \cite{we2}, the calculations were done using eq. (\ref{4}) and smooth
model dependent $T$-matrix solutions which could have possibly
missed or shifted some peaks. 
As in the standard baryon case, here too we see  
small structures in phase shift which give rise to time delay peaks.
We see in Fig. 2, in the $D_{03}$ partial wave, a peak at
$1.545$ GeV, followed by other higher lying resonances. In \cite{we2} this
corresponds to the value of $1.49$ GeV. In Fig. 3 (the $P_{01}$ partial
wave), we find the lowest resonance at $1.6$ GeV which corresponds 
in \cite{we2} to a peak at $1.57$ GeV. Note that in hadronic resonances, 
the resonance parameters quoted by different groups, often 
differ from each other (by some $10-20\%$) depending on 
the experiment and method used \cite{pdg}. We therefore conclude that the 
values we found here and in
\cite{we2}, are in good agreement with the recent experiments 
\cite{nakano,clas}.

It was recently pointed out \cite{close} that if the $\Theta$ (1540) is
$udud\bar{s}$ with $J^p=1/2^+$, then the correlations among QCD forces
necessarily imply the existence of $\Theta^*$ with $J^p=3/2^+$ which
is probably only slightly more massive than the $\Theta$ 
(the mass difference between $\Theta$
and $\Theta^*$ would be decided by the strength of the spin-orbit
forces within this exotic). The spin-orbit partner with $J=3/2$ 
is also predicted within a chiral constituent quark model in \cite{gloz},  
in contrast to the soliton model which rules out the existence of 
such a partner. This possible $\Theta^*$ is expected to be broad \cite{jenn}. 
In view of these predictions, we found it important to analyse the 
time delay in the $P_{03}$ partial wave where a possible $J=3/2$ partner of
the resonance at 1.6 GeV in the $P_{01}$ ($J=1/2$) partial wave could exist.
The $J=1/2$ partner of the $D_{03}$ resonance at 1.545 GeV (Fig. 2) would
correspond to the $D_{01}$ ($J=1/2$) partial wave, 
the data on which does not exist in the case of $K^+N$ elastic scattering 
(since $S=1/2$ for the $K^+N$ system, $L=2$ allows only $J=3/2,5/2$). 
In Fig. 4 we show the time delay analysis for $P_{03}$ and indeed find
similar peaks (around 1.6 and 1.8 GeV) as in the $P_{01}$ case.
However, since the error bars on the phase shifts in the $P_{03}$ case
are large, this finding of the $J=3/2$ partner should be treated with caution. 
Note also that the model solution (dashed line) for phase shift would 
not give rise to any resonant structure. 
We do think, however, that such an agreement between peak positions in
the $P_{01}$ and $P_{03}$ partial waves cannot be a coincidence and 
the spin-orbit partner will
most likely be confirmed when better data on $K^+N$ elastic scattering 
becomes available.

The widths of the resonances found in our analysis using time delay seem
to be somewhat larger than those predicted by the recent experiments
\cite{nakano,clas}. However, two of these experiments have been done
on nuclei, using carbon and xenon, which could lead to a reduction of
the resonance width as compared to that in free space. 
The width inside nuclear medium could reduce due to (a) modification
of the pentaquark-kaon-nucleon vertex inside the medium and (b) interaction 
of the decay products with the nucleus, which involves Pauli blocking
of nucleon states as well as modification of the $K^+$ propagator. 
Studies of similar effects in other resonances \cite{width}, have 
indeed shown the nuclear effects to be important.  
     
The above interpretation could however have a caveat worth discussing.
One would not expect any inelasticities to be present in the 
the relevant $1.5$ GeV region due to its being  close to the $KN$ threshold.  
Hence, as obvious from Fig. 1 
for the prominent $D_{13}$ resonance, the phase shift is expected to jump 
by $\pi$. This does not happen in Figs 2-4. The reason behind 
it could be twofold. Either the positive time delay peak in the $1.5$ 
GeV region does
not signify a proper resonance, or as explained below, a strong non-resonant 
background exists in this region. The first possibility casts strong 
doubts on interpreting
the experimentally found bump in the cross section as a proper resonance simply
because the phase shifts in any of the partial waves in the same 
energy region fail to make a `$\pi$-jump'.
Indeed, not every peak in the cross section is necessarily a resonance 
\cite{ginzburg}. Such an interpretation would be in favour of the fact
that no pole values have been found in the $1.5$ GeV region through partial
wave analysis and also in agreement with the conclusion in \cite{krein}.
The second possibility of a non-resonant background cannot be excluded 
a priori. Already in potential scattering, the total phase shift is a sum
of the resonant and non-resonant part. The latter is usually written as
$\tan^{-1}[j_l(ka)/n_l(ka)]$ where $k$ is the momentum, $a$ the 
potential range and $j_l$, $n_l$ are linear combinations of spherical
Hankel functions \cite{joachain}. Certainly, 
the aforementioned `$\pi$-jump' 
occurs only completely if the non-resonant part is missing. To avoid the
reference to potential scattering one might also parametrize the $S$ matrix
with the inclusion of the energy dependent background phase 
$\eta_i$ \cite{background} as
\begin{equation}
S = e^{2i\eta_i} \biggl [ 1 - 2i {E_R \Gamma_i(E) \over E^2 - E_R^2 + 
i E_R \Gamma(E)} \biggr ]
\end{equation}
which again leads to a distortion of the phase shift and hence time delay too. 
If we accept the peak around $1.5$ GeV in the cross section and the time delay
as a proper resonance, there seems to be no way out as to accept also
a strong non-resonant background. To decide which one of the two possibilities
is the most likely one, we give two physical examples of phase
shifts in strong interaction. The $P_{1/2}$ phase shift in $p-\alpha$
elastic scattering jumps from $0^0$ to about $60^0$ giving
the first excited $^5Li$ state which lies much below the first inelastic
threshold, namely the $p + \alpha \rightarrow d + \,^3He$ reaction 
\cite{arndtx}.
Another example is that of the $P_{1/2}$ level of $^5He$ found in
elastic $n-\alpha$ scattering where again the phase shift jumps
only by about $40^0$ \cite{morgan}. In both cases the phase shifts are
neither steep nor do they perform the `$\pi$-jump' due to the
presence of hard sphere potential scattering (non-resonant background
in other words). Note however that the above resonances are well
established.

The experimental papers do not give
the $L_{2I,2J}$ assignments of the resonance around 1.5 GeV. 
If we attribute the form of the phase shift due to a resonant and 
background part, we can 
speculate from our time delay plots (Figs 2-4) and our earlier
work \cite{we2}, that this resonance could either be $D_{03}$, $P_{01}$ 
or $P_{03}$.    
As mentioned in \cite{diak}, we would also like to note that the
older $K^+N$ scattering data is available from around 1.525 GeV onwards 
centre of mass energy of the $KN$ system. This could be the possible
reason as to why the low-lying exotic was not spotted by others in the
past. However, the time delay method clearly displays the peaks in the
$D_{03}$ and $P_{01}$ partial waves at 1.545 and 1.6 GeV respectively.   
Though not with very reliable data, we do find peaks in the $P_{03}$ 
partial wave which could be the $J=3/2$ partners of the ones in $P_{01}$. 

Given the success of the time delay method using phase shifts \cite{we1}, 
and the fact that we do find the 
recently reported exotic resonance around 1.5 GeV, we think
that the higher lying resonances in our time delay plots should also be
taken seriously. 
More so as the argument concerning the small inelasticity at
threshold is not valid anymore. It would be useful to improve the statistics
of the present experiments \cite{nakano,clas} in the energy region of 
1.8 GeV, to confirm the higher exotic states. 
\vskip1cm
{\bf Acknowledgement}\\
We wish to thank B. K. Jain for useful discussions. We also wish to thank
the anonymous referee for useful comments.

\end{document}